# Magnéli-Phases in Anatase Strongly Promote Co-Catalyst-Free Photocatalytic Hydrogen Evolution

Maximilian Domaschke,[#a] Xuemei Zhou,[#b] Lukas Wergen,[a] Stefan Romeis,[a]

Matthias E. Miehlich,[c] Karsten Meyer,[c] Wolfgang Peukert,*[a] Patrik Schmuki*[b]

a. Institute of Particle Technology, University of Erlangen-Nuremberg, Cauerstr. 4, 91058 Erlangen, Germany.

b. Department of Materials Science WW-4 LKO, University of Erlangen-Nuremberg, Martensstr. 7, 91058 Erlangen, Germany.

c. Department of Chemistry and Pharmacy, Inorganic & General Chemistry, University of Erlangen-Nuremberg, Egerlandstrasse 1, 91058 Erlangen, Germany.

ABSTRACT

Magnéli phases of titanium dioxide (such as $Ti_4O_7$, $Ti_5O_9$, etc.) provide electronic properties, namely a stable metallic behavior at room temperature. In this manuscript, we demonstrate that nanoscopic Magnéli phases, formed intrinsically in anatase during a thermal aerosol synthesis, can enable significant photocatalytic $H_2$ generation – this without the use of any extrinsic co-catalyst in anatase. Under optimized conditions, mixed phase particles of 30% anatase, 25% $Ti_4O_7$ and 20% $Ti_5O_9$ are obtained that can provide, under solar light, direct photocatalytic $H_2$ evolution at a rate of 145 µmol h$^{-1}$ g$^{-1}$. These anatase particles contain 5-10 nm size inter-grown phases of $Ti_4O_7$ and $Ti_5O_9$. Key is the metallic band of $Ti_4O_7$ that induces a particle internal charge separation and transfer cascade with suitable energetics and favorable dimensions that are highly effective for $H_2$ generation.

Magnéli phases are a set of titanium sub-oxides with a general composition of $Ti_nO_{2n-1}$, where n varies from 3 to 10, such as $Ti_4O_7$, $Ti_5O_9$, etc. The structures of this specific set of titanium suboxides were described in the 1950s by Magnéli et al..[1-4] Typical synthesis routes involve carbothermal reduction,[5] reducing rutile with $CaH_2$ at 350 °C,[6] or reducing anatase or rutile in hydrogen atmosphere at temperatures typically >1200 °C.[7] The compounds have in common that they consist of distorted $TiO_6$ octahedra which are joined by sharing edges to form an infinite three-dimensional framework.[6] The deviation from the ideal $TiO_2$ stoichiometry is accommodated for by oxygen vacancies regularly arranged throughout the lattice.[8-11] As a result, many Magnéli phases provide an electronic structure that shows a metallic behavior at room temperatures.[12-16] For example, for $Ti_4O_7$ the Fermi level at room temperature is located within the oxide's conduction band and the material has an electronic conductivity that is comparable to metals.[17-19] Due to this feature, Magnéli phases are industrially used as conductive oxide electrodes that provide a wide electrochemical window, comparable or superior to graphite or lead.[20, 21]

Interestingly, theoretical work also shows the filled metallic band of the Magnéli phases, namely $Ti_4O_7$, to be located close to the conduction band edge of $TiO_2$. This makes Magnéli phases interesting in microelectronics and in particular in memristic devices where semiconductor/metal switching is exploited.[5, 7, 22-27] Nevertheless, such semiconductor/metal junctions are also of a high interest for photocatalysis, where efficient charge separation within a catalyst particle and transfer of the photoinduced charge carriers to the surrounding media dictates the efficiency of a photocatalyst. Also in the case of anatase $TiO_2$ - the most widely investigated economic and stable photocatalyst with an energetics that allows water oxidation and reduction - kinetic limitations of the reactions can be overcome by designing suitable electronic junctions or defect structures.[28, 29] Energetic gradients in junctions can give charge carriers separation directionality and lower the kinetic surface barriers, as in the case of noble metal co-catalysts (Au, Pt, Pd) that are widely used, when $H_2$ generation from $TiO_2$ should be achieved at reasonable rates.

For the effectiveness of a particle internal junction it is of key importance that the dimensions are adjusted to the carrier diffusion/drift length. With a typical hole diffusion length in the range of some 10 nm in anatase[30], size and energy matching junction engineering is desired. For example Bahnemann et al. discuss a variety of critical processing factors that affect the $H_2$ evolution performance when anatase/rutile photocatalysts are formed by wet synthesis.[31] Previous work also used several reduction treatments to create low concentration $Ti^{3+}$-Ov states that were able to mediate electron transfer from anatase particles.[32] In contrast in the present work, we use a $H_2$ stream thermal aerosol hot-wall reactor synthesis that is able to induce nanoscopic Magnéli phase segments into growing anatase nanocrystals. The synthesis is described in detail in the SI and allows to fabricate $Ti_4O_7$ and $Ti_5O_9$ containing $TiO_2$ nanoparticles with a narrow size distribution (particle size <100 nm) and with a variable ratio of anatase to Magnéli phase. We find these particles to exhibit the remarkable feature to strongly enhance photocatalytic $H_2$ generation without the use of any external co-catalyst.

**Figure 1a** shows XRD spectra for samples synthesized at different processing temperatures (from 500 °C to 1000 °C) and **Figure 1b** gives the evaluation of the phase composition of the powder. Rietveld analysis of the diffractograms shows that at temperatures of up to 500 °C only anatase is formed. At higher temperatures (800 °C), small amounts of Magnéli phase (< 5%) start to appear. For 900 °C, clear peaks for $Ti_3O_5$ (PDF No. 01-072-2101, $Ti_3O_5$, Orthorhombic, Cmcm space group), $Ti_4O_7$ (PDF No. 01-072-1724, $Ti_4O_7$, Anorthic, A-1 space group), $Ti_5O_9$ (PDF No. 01-071-0627, $Ti_5O_9$, Anorthic, P-1 space group) can be observed, with a phase composition of 30% anatase, 15% rutile and 55% Magnéli phase. At 1000 °C, Magnéli phases reach 60% $Ti_4O_7$ and 20% $Ti_5O_9$ and < 10% anatase and rutile are present. For samples prepared at 1100 °C, XRD shows $Ti_4O_7$ to be the main crystalline phase in the material with only < 5% anatase left. Importantly, **Figure S1** shows that our synthesis approach leads only to a slight increase of the particle size from 800 °C to 1100 °C, that is from approx. 22 nm at 800 °C to approx. 49 nm at 1100 °C.

Changes in the phase composition could be confirmed by Raman measurements (**Figure S4**). The inset in **Figure 1c** shows that the conversion to particles with an increased sub-oxide content is also accompanied with a color change of the powder from white to dark blue. The samples maintain this appearance for several months at ambient conditions and at temperatures of up to 80 °C in air.

These powders were then investigated for their photocatalytic $H_2$ evolution activity under AM 1.5 (100 mW cm$^{-2}$) solar simulator illumination and ultra violet light ($\lambda$= 365 nm, 90 mW cm$^{-2}$), using suspensions of the particles in a $H_2O$/methanol solution. Please note that all experiments were performed without the use of noble metal co-catalyst (experimental details are given in the Supporting Information). The results in **Figure 1c** and **Figure S2** show that titania particles become increasingly active for $H_2$ evolution with increasing temperature under plain UV and solar irradiation up to 900 °C. An optimum activity is obtained for powder synthesized at 900 °C, a photocatalytic $H_2$ evolution rate of ≈145 µmol h$^{-1}$ g$^{-1}$ is achieved.

For higher synthesis temperatures (>1000 °C), the materials show a diminishing activity, and for the sample prepared at 1100 °C, no $H_2$ evolution is detectable. This is in line with experiments performed with an untreated commercial Magnéli powder that was used for reference. It is noteworthy that a mild contribution from visible light to the overall $H_2$ production can be observed for the Magnéli-phase containing powder produced at 900 °C (**Figure S2b**). For this sample also repeated photocatalytic tests in cycling experiments (**Figure S2a**) show that the $H_2$ generation activity is stable over the investigated time interval of 24 h.

**Figure 1d** shows a high resolution HRTEM image of the most active sample, i.e. the sample prepared at 900 °C. The particle essentially consists of adjacent phases of $Ti_4O_7$ and anatase. The observed lattice fringes correspond to anatase [101] with a d spacing of 0.351 nm (i.e., corresponding to the XRD peak at 2θ = 25.2°) and a spacing for $Ti_4O_7$ of 0.331 nm, which corresponds to the [1$\bar{2}$0] facet of $Ti_4O_7$. I.e., the most active sample consists essentially of intergrown phases of anatase and $Ti_4O_7$ with similar domain sizes of 5-10 nm (the entire particle is approx. 30 nm in diameter). Also, from the SAED pattern the typical features of anatase and $Ti_4O_7$ can be identified but additionally traces of rutile [111] can be identified (**Figure 1d**). HRTEM images for a sample synthesized at 1000 °C (**Figure S3**) shows lattice spacings of 0.260 nm, 0.409 nm and 0.246 nm, which can be ascribed to the [200] facet of $Ti_4O_7$, the [$\bar{1}$02] facet of $Ti_5O_9$ and the [$\bar{1}\bar{2}$1] of $Ti_5O_9$ and [120] of $Ti_4O_7$. This is in line with XRD that indicate that the sample prepared at 1000 °C contains only 5% anatase, 20% $Ti_5O_9$ and 60% $Ti_4O_7$.

X-ray photoelectron spectroscopy (XPS) data are shown in **Figure 2a** for the O1s peaks and in **Figure 2b** for the Ti2p peaks of samples prepared at 500 °C, 900 °C and 1100 °C. The O1s peak can be deconvoluted into three contributing peaks. The main peak at 529.6 eV corresponds to oxygen in the $TiO_2$ position. This peak intensity decreases with higher synthesis temperature (from 75.3% at 500 °C to 48.1% at 1100 °C). The peak at 531.9 eV can be attributed to defective oxygen species from Magnéli phase as described in literature[6, 33, 34] and further confirmed by spectra taken on reference commercial $Ti_4O_7$ (**Figure S5**). This peak increases from 19.0% at 500 °C to 34.0% at 1100 °C.

The concentration at 1100 °C of 34.0% is well in line with the reference $Ti_4O_7$ (**Figure S5**), where a contribution of 36.3% of this peak to the overall spectrum is obtained. The peak at higher binding energy (533.6 eV) can be ascribed to surface adsorbed water. [35] The Ti2p peak for commercial Magnéli phase (**Figure S5**) shows two clear peaks at a binding energy of 456.0 eV and 461.7 eV, which can be attributed to titanium in nonstoichiometric $TiO_2$.[36] The Ti2p peaks for sample prepared at 1100 °C (**Figure 2b**) shows only a mild sub-stoichiometry contribution, which may be ascribed to the remaining anatase in the powder. However, overall, these findings are well in line with the formation of sub-oxides in the $TiO_2$ as well as the commercial Magnéli phase. [37]

**Figure 2c** shows CW X-band EPR spectra for samples prepared in a temperature range from 500 °C to 1100 °C recorded at 100 K. The spectrum of the pure anatase (sample prepared at 500 °C), shows trace amounts of characteristic intrinsic defects ("F-centers" or oxygen vacancies in anatase [38]) at g≈ 2.0. For samples synthesized at high temperature (800 °C to 1100 °C), and where XRD indicates the presence of a Magnéli phase, a broader and significantly more intense signal is apparent with a g-values ranging from 1.95-1.96. Compared with the EPR spectra of commercially available Magnéli crystallites (**Figure S6**) and literature data,[34] this signal is attributed to a $(Ti^{3+}V_o\text{-}Ti^{4+})^{+1}$ center, which consists of an electron localized on or about an oxygen vacancy with a possible charge transfer between the two Ti sites.[39]

To obtain information on the electronic properties of the Magnéli phase containing samples, we carried out photoelectrochemical measurements as shown in **Figure 3a** and **Figure 3b**. Photocurrent spectra were taken from powder film samples on FTO for samples prepared at 500 °C, 800 °C, 900 °C and 1100 °C, and a commercial Magnéli phase pellet in 0.1 M $Na_2SO_4$ (see SI for details).

**Figure 3a** shows that with the increase of synthesis temperature, a decrease of the magnitude of the photocurrent in the UV is observed. The sample prepared at 1100 °C shows the clearly lowest photocurrent response but a very noticeable tail into the visible region. The band gap evaluation in **Figure 3b** gives a value of 3.21 eV for sample prepared at 500 °C, which corresponds well with values for the indirect band gap of anatase. However, for samples prepared at 800 °C and 900 °C, the band gap evaluation leads to an increasingly lower value of 3.15 eV for 800 °C 3.00 eV for 900 °C. This shift can be ascribed to the increasing contribution of sub-oxide species with occupied states close to the conduction band of anatase.[40, 41]

To explain the beneficial effect of an anatase/$Ti_4O_7$ junction on the photocatalytic hydrogen generation one may consider previous theoretical work on the electronic structure of Magnéli phases and namely $Ti_4O_7$.[37, 41] At room temperature, DOS calculations for $Ti_4O_7$ show a metallic band of a width of approx. 0.5 eV that consists of 3d Ti states with a formal $Ti^{+3.5}$ charge. The electron hopping activation energies to move an electron from a $Ti^{3+}$ to a neighboring $Ti^{4+}$ is < 0.2 eV (i.e. sufficient to provide room temperature conductivity). The Fermi level of this metallic band, in absolute energy, is located approx. 0.1 – 0.2 eV below the conduction band of anatase. The energetic situation of an anatase/$Ti_4O_7$ junction can thus roughly be outlined as shown in **Figure 3c**. From these considerations, one can deduce $Ti_4O_7$ to act as a charge transfer mediator for photogenerated electrons (similar to a noble metal co-catalyst). Anatase, to a large extent, acts as the carrier harvesting phase whereas the $Ti_4O_7$ extracts electrons and mediates their transfer to the surrounding liquid phase. Due to the close proximity of anatase and Magnéli phase (in a single particle), charge carrier separation can be very efficient. Additionally, the Magnéli phase provides a high electronic conductivity, which ensures a high carrier mobility through the co-catalyst and thus further facilitates the charge separation in the photocatalytic process.

In conclusion, the present work shows that nanoscopic inserts of Magnéli phases, formed directly during anatase titania nanoparticle synthesis, can act as efficient co-catalyst in anatase for photocatalytic $H_2$ generation. For particle sizes <100 nm, anatase/$Ti_4O_7$ internal junctions are highly beneficial for charge separation, as with a size of ≈ 5-10 nm, the junction dimensions lie within the diffusion length of holes in anatase.[42] Mixed phase particles of 30% anatase, 25% $Ti_4O_7$ and 20% $Ti_5O_9$ show the highest photocatalytic $H_2$ evolution activity (145 µmol h$^{-1}$ g$^{-1}$), without the use of any extrinsic co-catalyst. Pure anatase or pure Magnéli phase yield significantly weaker photocatalytic $H_2$ generation activity. We attribute the enhanced photocatalytic performance of these mixed particles to the electronic properties of the Magnéli phases, where anatase acts as light harvester and $Ti_4O_7$ acts as mediator for charge separation, as well as for electron transport and transfer.[43] Furthermore, the synthesis approach used here allows for a wide in-situ variation of features, such as doping and decoration, and provides a potent tool for further direct modification of Magnéli/anatase based photocatalysts.

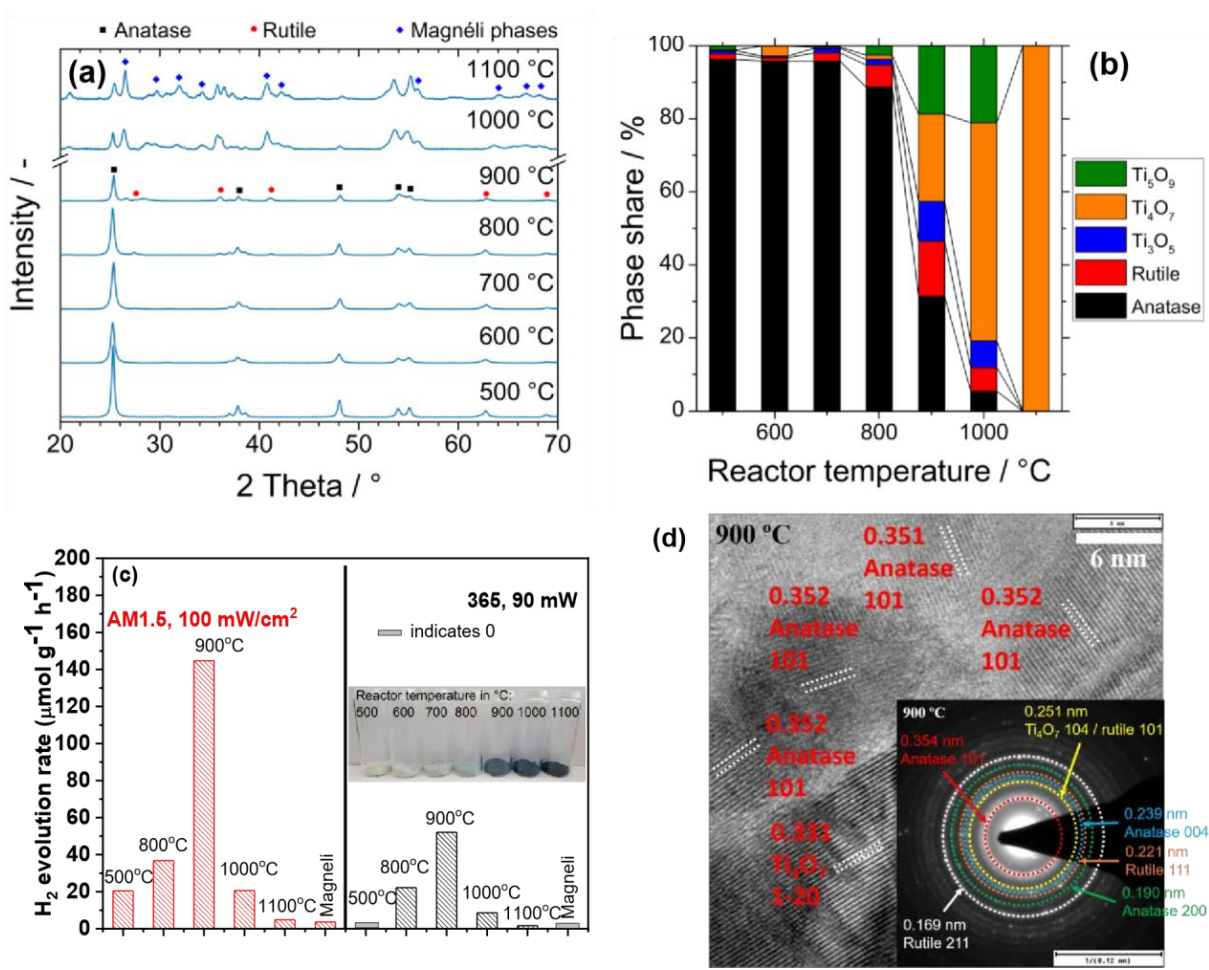

**Figure 1**. (a) XRD patterns for the as-synthesized materials. (b) Phase composition analysis based on **Figure 1a**. (c) Photocatalytic $H_2$ evolution from different phase coupled titania corresponding

to Figure 1a under 365 nm LED or AM1.5 (100 mW cm$^{-2}$) solar simulator illumination measured in a 50 vol% methanol–water electrolyte. Inset: optical images of the samples. (d). HRTEM images for sample prepared at 900 °C. Inset: SAED pattern and its analysis for the sample with maximum H$_2$ evolution activity (prepared at 900 °C.)

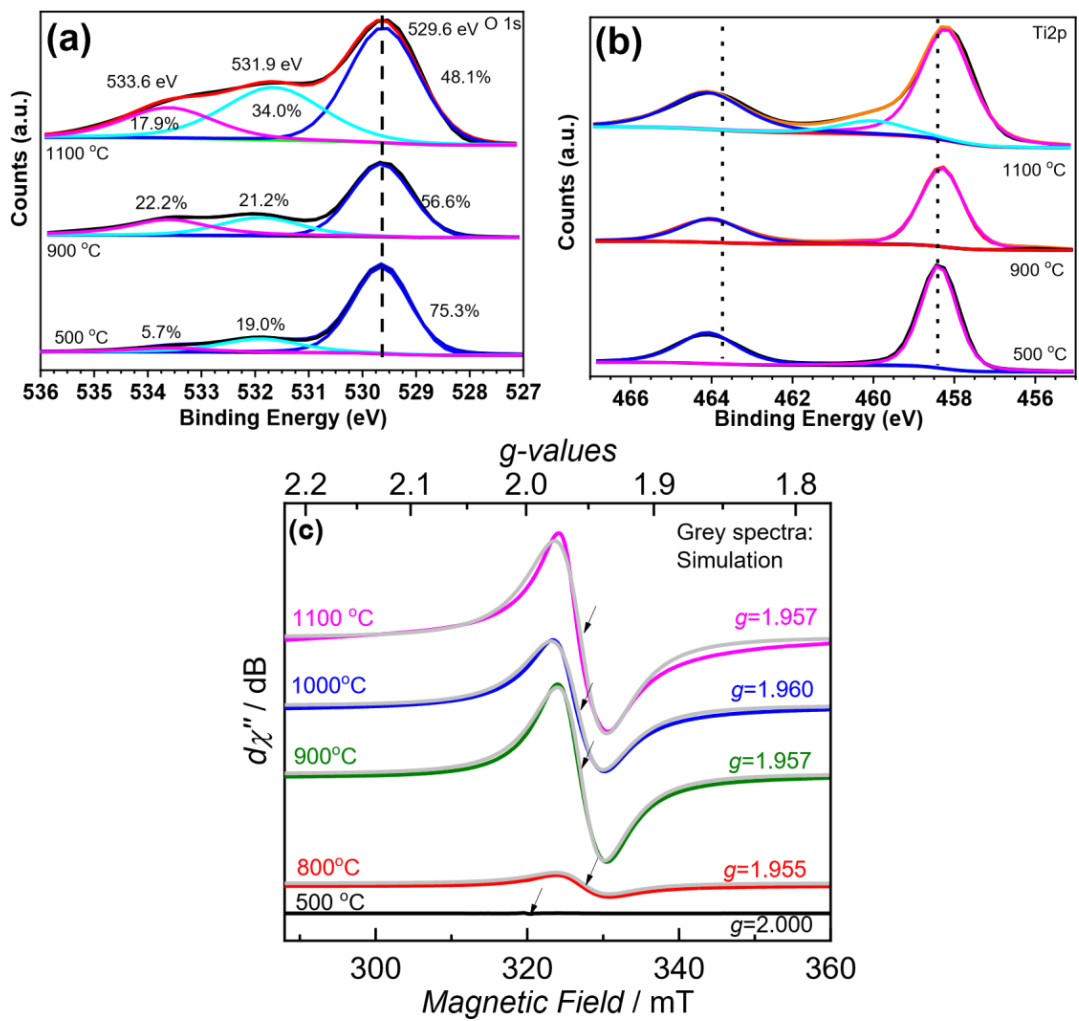

**Figure 2**. High resolution XPS (a) Ti2p and (b) O1s peaks for samples prepared at 500 °C, 900 °C and 1100 °C. (c) Solid-state EPR spectra for materials prepared from 500 °C to 1100 °C, measured at 100 K.

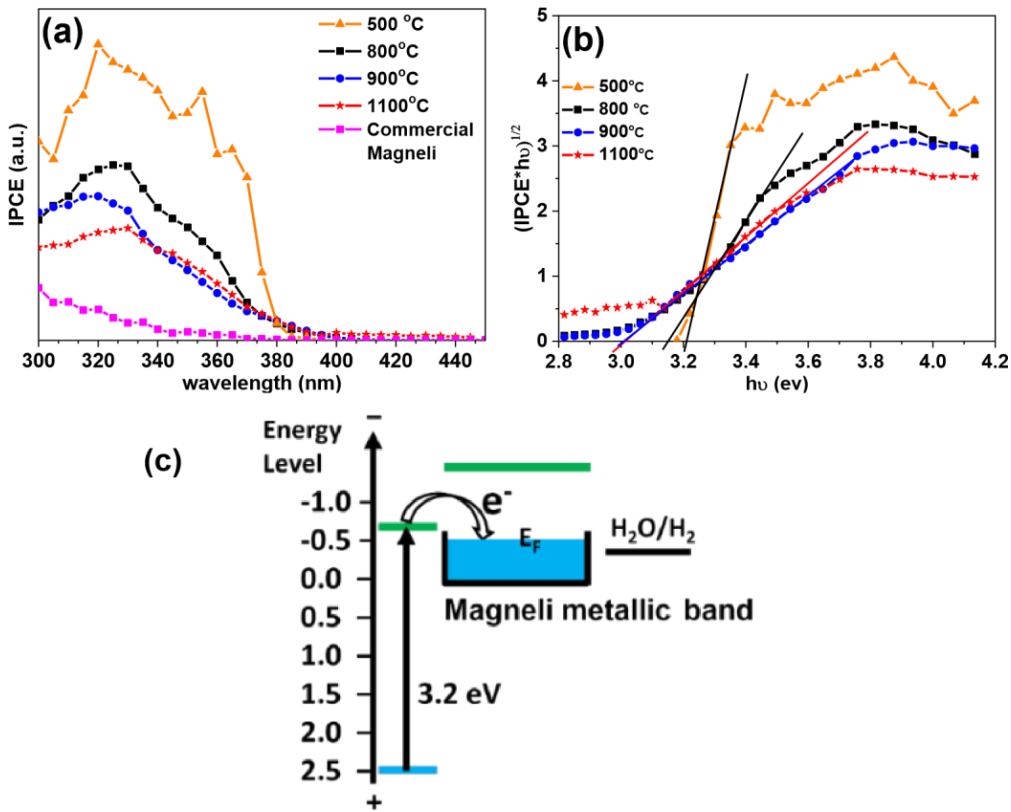

**Figure 3** (a) Incident photocurrent conversion efficiency (IPCE) measurements for the assynthesized materials as electrodes and (b) band gap evaluation from spectra in (a). (c) Schematic drawing for the band alignment between TiO$_2$ and Magnéli phase based on theory.

**Corresponding Author**

Wolfgang Peukert: E-mail: wolfgang.peukert@fau.de;

Patrik Schmuki: E-mail: schmuki@ww.uni-erlangen.de.

**Author Contributions**

The manuscript was written through contributions of all authors. All authors have given approval to the final version of the manuscript. # indicates equal contribution of the work by the authors.

**Conflicts of interest**

There are no conflicts to declare.

**Supporting Information**

Supporting Information include experimental section, SEM images of particles, stability tests of photocatalytic activity, Raman spectra, TEM images, XPS peaks and EPR of commercial Magnéli powder.

**Acknowledgement**

Financial support from ERC and DFG within the framework of its Excellence Initiative for the Cluster of Excellence "Engineering of Advanced Materials" is thankfully acknowledged.

Table of Contents

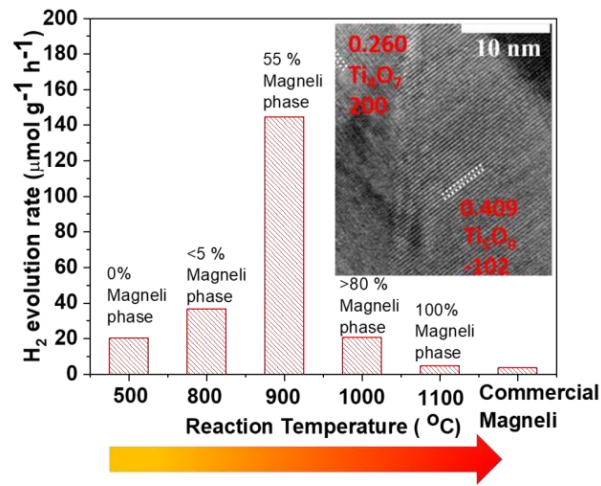

Nanoscopic Magnéli phases, formed intrinsically in anatase, can enable significant photocatalytic H$_2$ generation – this without the use of any extrinsic co-catalyst in anatase. Key is the metallic band of Ti$_4$O$_7$ that induces a particle internal charge separation and transfer cascade with suitable energetics and favorable dimensions that are highly effective for H$_2$ generation.